\documentclass[aps,showpacs,preprintnumbers,amsmath,amssymb]{revtex4}
\usepackage{dcolumn}
\usepackage[dvips]{epsfig}
\usepackage{graphicx}
\usepackage{amssymb}
\usepackage{color}
\usepackage{enumerate}
\usepackage{subfigure}

\begin{document}
\baselineskip=0.8 cm
\title{Effect of gravitational wave on shadow of a Schwarzschild black hole}
\author{Mingzhi Wang$^{1}$\footnote{Corresponding author: wmz9085@126.com},  Songbai Chen$^{2,3}$\footnote{csb3752@hunnu.edu.cn},  Jiliang
Jing$^{2,3}$\footnote{jljing@hunnu.edu.cn}
}
\affiliation{$ ^1$School of Mathematics and Physics, Qingdao University of Science and Technology, Qingdao, Shandong 266061, People's Republic of
China \\ $ ^2$Institute of Physics and Department of Physics, Key Laboratory of Low Dimensional Quantum Structures
and Quantum Control of Ministry of Education, Synergetic Innovation Center for Quantum Effects and Applications,
Hunan Normal University, Changsha, Hunan 410081, People's Republic
of China\\
$ ^3$Center for Gravitation and Cosmology, College of Physical Science and Technology,
Yangzhou University, Yangzhou 225009, China}

\begin{abstract}
\baselineskip=0.6 cm
\begin{center}
{\bf Abstract}
\end{center}

We have studied the shadows of a Schwarzschild black hole under a special polar gravitational perturbation, which is a particular solution of Einstein equations expanded up to first order. It is shown that the black hole shadow changes periodically with time and the change of shadow depends on the Legendre polynomial order parameter $l$ and the frequency $\sigma$ of gravitational wave. For the odd order of Legendre polynomial, the center of shadow oscillates along the direction which is vertical to equatorial plane. For even $l$, the center of shadow does not move, but the shadow alternately stretches and squeezes with time along the vertical direction. Moreover, the presence of the gravitational wave leads to the self-similar fractal structures appearing in the boundary of the black hole shadow. We also find that this special gravitational wave has a greater influence on the vertical direction of black hole shadow.

\end{abstract}
\maketitle

\newpage
\section{Introduction}
The first direct observation of gravitational waves (GW150914)\cite{js19,js20,js21} reported by LIGO and Virgo Scientific not only confirms the success of Einstein's general relativity, but also opens a new era in the fields of astronomy, astrophysics and cosmology. Subsequently, there are several gravitational waves events have been detected, which are caused by binary black hole merger  \cite{js19,js20,js21,gw1,gw2,gw3,gw4} or by binary neutron star  merger \cite{gw5}. Especially, the discovery of the electromagnetic signals in gamma-ray \cite{gw5,sub3,sub4} arising from  binary neutron star (BNS) merger means the arrival of multi-messengers astronomy. By comparing with theoretical templates, gravitational waves could tell us a variety of parameters of astrophysical compact objects such as their masses, spins and so on. The detection of gravitational waves could help us to understand black hole further and to verify various gravity theories.

Another exciting event in astrophysics and black hole physics is the first image of the supermassive black hole in the center of the giant elliptical galaxy M87, which was announced by Event Horizon Telescope (EHT) Collaboration in 2019 \cite{eht,fbhs1,fbhs2,fbhs3,fbhs4,fbhs5,fbhs6}. It provides the first direct visual evidence that there exists exactly black hole in our Universe. Black hole image can be regarded as a potential tool to verify  gravity theories and identify black hole parameters. The initial analyses of the first image of black hole have no striking deviations from the predictions of general theory of relativity. The dark region in the center of black hole image is black hole shadow, which corresponds to light rays fall into event horizon of black hole. The fingerprints of the geometry around the black hole would be reflected in the shape and size of black hole shadow\cite{sha2,sha3}. For example, the shadow of a Schwarzschild black hole is a perfect black disk. But for a Kerr black hole, shadow becomes a D-shaped silhouette gradually with the increase of spin parameter\cite{sha2,sha3}.  In the spacetime of a Kerr black hole with Proca hair and a Konoplya-Zhidenko rotating non-Kerr black hole, the cusp silhouette of black hole shadows emerge in certain range of parameter \cite{fpos2,sb10}. Especially,  the self-similar fractal structures are found in shadow for a rotating black hole with scalar hair \cite{sw,swo,astro,chaotic}, a Majumdar-Papapetrou binary black hole system \cite{binary, sha18}, Bonnor black diholes with magnetic dipole moment \cite{my}, and a non-Kerr rotating compact object with quadrupole mass moment \cite{sMN}. These novel structures in shadows are found to be caused by non-integrable photon motions. The black hole shadows with other parameters in various theories of gravity have been recently investigated in Refs. \cite{swo7,mbw,sha4,sha5,sha6,sha7,sha8,sha9,sha10,sha11,sha111,sha12,sha13,sha14,sha141,sha15,sha16,
sb1,sha17,sha19,sha191,sha192,sha193,sha194,shan1,shan1add,shan2add,shan3add,drk,rr,pe,lf2,Zeng2020vsj,Zeng2020dco}. These could provide some theoretical templates for the future astronomical observations announced by Event Horizon Telescope and BlackHoleCam\cite{bhc}.

Since both the gravitational waves detection and Event Horizon Telescope observation play a vital role in the study of black holes and verification of various gravity theories, it is very interesting to study the effects of gravitational waves on black hole shadows. Generally, it is very difficult because that it is not easy to get a solution for the gravitational wave around a black hole since the Einstein equation with perturbation is very complicated. Fortunately, B. Xanthopoulos \cite{cgw16} obtained a particular gravitational wave solution which meets Einstein equations expanded up to the first order in $\epsilon$. With this particular solution, one can probe the peculiar effects of gravitational wave on dynamics of test particle in black hole spacetime.
It is found that this special gravitational wave makes the motion of a timelike test particle is no longer integrable and then the chaotic phenomenon appears \cite{cgw}, which is different from those in the case without gravitational wave. It is naturally expected that the chaotic phenomenon could appear in the motion of photon under such gravitational perturbation, and then the corresponding chaotic behavior of photon would give rise to some new effects on the black hole shadow. Therefore, in this paper, we would like to probe the effects of this gravitational wave on the black hole shadow.

The paper is organized as follows. In Sec. II, we review briefly the spacetime of a Schwarzschild black hole perturbed by the gravitational wave\cite{cgw16,cgw,cgw18} and then analyze the null geodesic equations in this spacetime. In Sec. III, we present numerically the shadows for the Schwarzschild black hole perturbed by the gravitational wave and probe the effects of this gravitational wave on the shadow. Finally, we present a summary.

\section{The spacetime of Schwarzschild black hole perturbed by gravitational wave and null geodesics}

The metric of a Schwarzschild black hole with a gravitational perturbation \cite{cgw16,cgw,cgw18} can be expressed as
\begin{eqnarray}\label{dgmtr}
ds^{2}=(g_{\mu\nu}+\epsilon h_{\mu\nu})dx^{\mu}dx^{\nu}.
\end{eqnarray}
which is a particular solution of Einstein equations expanded up to the first order in $\epsilon$. The metric $g_{\mu\nu}$ in Eq. (\ref{dgmtr}) is the metric tensor of a usual Schwarzschild black hole with a form
\begin{eqnarray}
g_{tt}=-f=-(1-2M/r),\quad \quad  g_{rr}=f^{-1},\quad \quad
g_{\theta\theta}=r^{2},\quad \quad
g_{\phi\phi}=r^{2}\sin^{2}\theta.
\end{eqnarray}
$h_{\mu\nu}$ is a special analytical solution of polar gravitational perturbation around Schwarzschild black hole \cite{cgw16,cgw,cgw18}, i.e.,
\begin{eqnarray}
\label{dg}
&&h_{tt}=-fXP_{l}\cos(\sigma t),\\ \nonumber
&&h_{rr}=f^{-1}YP_{l}\cos(\sigma t),\\ \nonumber
&&h_{\theta\theta}=r^{2}(ZP_{l}+W\frac{d^{2}P_{l}}{d\theta^{2}})\cos(\sigma t),\\ \nonumber
&&h_{\phi\phi}=r^{2}\sin^{2}\theta(ZP_{l}+W\frac{dP_{l}}{d\theta}\cot\theta)\cos(\sigma t),
\end{eqnarray}
where
\begin{eqnarray}
\label{dgx}
&&X=pq,\;\;\;\;\;\;\;\;\;\;\;\;\;\;\;Y=3Mq,\;\;\;\;\;\;\;\;\;\;\;\;\;\;\;Z=(r-3M)q,\;\;\;\;\;\;\;\;\;\;\;\;\;\;\;W=rq,\\ \nonumber &&p=M-\frac{M^{2}+\sigma^{2}r^{4}}{r-2M}, \;\;\;\;\;\;\;\;\;\;\;\;\;\;\;\;\;\;\;\;\;\;\;\;\;\;\;\;\;q=\frac{\sqrt{f}}{r^{2}}.
\end{eqnarray}
$P_{l}=P_{l}(\cos\theta)$ are the usual Legendre polynomials ($l>1$), $\sigma$ is the frequency of gravitational wave. This solution of the special class of gravitational wave (\ref{dg}) was obtained by B. Xanthopoulos \cite{cgw16} through solving the differential equations on metric perturbations of Reissner-Nordstr\"{o}mblack hole \cite{zer,mona, monb}. And the perturbations of Reissner-Nordstr$\ddot{o}$m black hole was derived by Zerilli \cite{zer} and Moncrief \cite{mona, monb} when they researched the perturbed gravitational and electromagnetic fields produced by a charged (or uncharged) test particle moving in a Reissner-Nordstr$\ddot{o}$m geometry. So the special class of gravitational wave (\ref{dg}) is a solution of one-dimensional wave-type equations\cite{zer,mona, monb}, and could be regarded as a perturbation caused by an uncharged particle moving in Schwarzschild black hole spacetime. The unperturbed (Schwarzschild) metric is spherically symmetric, thus there is no loss of generality that the special class of gravitational wave (\ref{dg}) is only an axisymmetric perturbation. Since $h_{\mu\nu}$ is divergent at infinity, the gravitational wave (\ref{dg}) is just used to describe the gravitational perturbation around a black hole. Actually, this solution is related to Zerilli function $Z^{(+)}$ by \cite{cgw16}
\begin{eqnarray}
Z^{(+)}=\frac{r^2}{3M+nr}(\frac{3MW}{r}-Y),
\end{eqnarray}
with $n=(l-1)(l+2)/2$.
For the even order $l$ in $P_{l}$, one can find that the gravitational wave solution (\ref{dg}) is even function of $\cos\theta$, therefore it is symmetric with respect equatorial plane \cite{cgw16,cgw,cgw18}. However, for the odd $l$, the perturbation (\ref{dg}) is odd function of $\cos\theta$ and then it is not symmetric with respect equatorial plane.

The Hamiltonian of a photon propagation along null geodesic in the spacetime (\ref{dgmtr}) can be expressed as
\begin{eqnarray}
\label{gwhami}
H=&&-\frac{p_{t}^{2}}{2f[1+\epsilon XP_{l}\cos(\sigma t)]}+\frac{fp_{r}^{2}}{2[1+\epsilon YP_{l}\cos(\sigma t)]}+\frac{p_{\theta}^{2}}{2r^{2}[1+\epsilon(ZP_{l}+W\frac{d^{2}P_{l}}{d\theta^{2}})\cos(\sigma t)]}\\ \nonumber &&+\frac{\csc^{2}\theta p_{\phi}^{2}}{2r^{2}[1+\epsilon(ZP_{l}+W\frac{dP_{l}}{d\theta}\cot\theta)\cos(\sigma t)]}.
\end{eqnarray}
It is obvious that the Hamiltonian (\ref{gwhami}) is an explicit function of time coordinate $t$, and then the energy of photon $E=-p_{t}$  \cite{fpos2, swo7,sha19,carter,rtcm,bhip} is no longer a constant of motion along geodesic ($\dot{E}=\dot{-p}_{t}=\frac{\partial H(t)}{\partial t}\neq0$), which can be expressed as
\begin{eqnarray}
\label{energy}
E=-p_{t}=f[1+\epsilon XP_{l}\cos(\sigma t)]\dot{t}.
\end{eqnarray}
It is understandable because there exists the energy transfer between the gravitational wave (\ref{dg}) and the photon in this spacetime. However, the $z$ component of the angular momentum $L_{z}=p_{\phi}$ of photon is still conserved in this case because the Hamiltonian (\ref{gwhami}) does not contain the coordinate $\phi$ \cite{fpos2, swo7,sha19,carter,rtcm,bhip}.
With the conserved quantity
\begin{eqnarray}
\label{L}
L_{z}=p_{\phi}=r^{2}\sin^{2}\theta[1+\epsilon(ZP_{l}+W\frac{dP_{l}}{d\theta}\cot\theta)\cos(\sigma t)]\dot{\phi},
\end{eqnarray}
we can find the null geodesic equations of photon in the spacetime (\ref{dgmtr}) can be expressed as
\begin{eqnarray}
\label{cdx}
\ddot{t}&&=\frac{1}{2 f \left[X \epsilon
   P_l \cos (\sigma  t)+1\right]}\bigg\{\frac{\dot{r}^2 \sigma  Y \epsilon  P_l \sin (\sigma  t)}{f}-2 \dot{r} \dot{t} \left[\epsilon  P_l \left(X f'+f X'\right) \cos (\sigma  t)+f'\right]-2 f \dot{\theta } \dot{t} X \epsilon
   \frac{dP_{l}}{d\theta} \cos (\sigma  t)\\ \nonumber &&+f \sigma  \dot{t}^2 X \epsilon  P_l \sin (\sigma  t)+\frac{L_{z}^2 \sigma  \epsilon  \csc ^2\theta \sin (\sigma  t) \left(W \cot\theta \frac{dP_{l}}{d\theta}+Z P_l\right)}{r^2
   \left[\epsilon  \cos (\sigma  t) \left(W \cot\theta  \frac{dP_{l}}{d\theta}+Z P_l\right)+1\right]{}^2}+\dot{\theta }^2 r^2 \sigma  \epsilon  \sin (\sigma  t) \left(W \frac{d^{2}P_{l}}{d\theta^{2}}+Z P_l\right)\bigg\},
\end{eqnarray}
\begin{eqnarray}
\label{cd1}
\ddot{r}&&=\frac{f}{2 \left[Y \epsilon  P_l\cos (\sigma  t)+1\right]} \bigg\{-\dot{t}^2 \left[\epsilon  P_l \left(X f'+f X'\right) \cos (\sigma  t)+f'\right]+\frac{\dot{r}^2 \left[\epsilon  P_l \left(Y f'-f Y'\right) \cos (\sigma  t)+f'\right]}{f^2}\\ \nonumber &&+\frac{2\dot{r} \sigma  \dot{t} Y \epsilon  P_l \sin (\sigma  t)}{f} +\frac{L_{z}^2 \csc ^2\theta\{ \epsilon  \cos(\sigma  t) \left[\cot\theta \left(r W'+2 W\right) \frac{dP_{l}}{d\theta}+P_l \left(r Z'+2 Z\right)\right]+2\}}{r^3 \left[\epsilon  \cos (\sigma  t) \left(W \cot\theta \frac{dP_{l}}{d\theta}+ZP_l\right)+1\right]{}^2}\\\nonumber &&+\dot{\theta }^2 r \bigg\{\epsilon  \cos (\sigma  t) \left[\left(r W'+2 W\right) \frac{d^{2}P_{l}}{d\theta^{2}}+P_l \left(r Z'+2 Z\right)\right]+2\bigg\}-\frac{2 \dot{\theta } \dot{r} Y \epsilon  \frac{dP_{l}}{d\theta} \cos (\sigma  t)}{f}\bigg\},
\end{eqnarray}
\begin{eqnarray}
\label{cd2}
\ddot{\theta}&&=\frac{1}{2 r^2 \left[\epsilon  \cos (\sigma  t) \left(W
   \frac{d^{2}P_{l}}{d\theta^{2}}+Z P_l\right)+1\right]}\bigg\{\frac{\dot{r}^2 Y \epsilon  \frac{dP_{l}}{d\theta} \cos (\sigma  t)}{f}+\dot{\theta }^2 r^2 \epsilon  \cos (\sigma  t) \left(W \frac{d^{3}P_{l}}{d\theta^{3}}+Z \frac{dP_{l}}{d\theta}\right)\\ \nonumber &&+\frac{L_{z}^2 \csc ^2\theta  \left\{2 \cot\theta+\epsilon  \cos (\sigma  t) \left[W
   \left(\cot ^2\theta-1\right) \frac{dP_{l}}{d\theta}+W \cot\theta \frac{d^{2}P_{l}}{d\theta^{2}}+Z \left(2 \cot\theta P_l+\frac{dP_{l}}{d\theta}\right)\right]\right\}}{r^2 \left[\epsilon  \cos (\sigma  t) \left(W \cot\theta \frac{dP_{l}}{d\theta}+Z
   P_l\right)+1\right]{}^2}\\ \nonumber &&+2 \dot{\theta } r^2 \sigma  \dot{t} \epsilon  \sin (\sigma  t) \left(W \frac{d^{2}P_{l}}{d\theta^{2}}+Z
   P_l\right)-2 \dot{\theta } \dot{r} r \left(\epsilon  \cos (\sigma  t) \left(\frac{d^{2}P_{l}}{d\theta^{2}} \left(r W'+2 W\right)+P_l \left(r Z'+2 Z\right)\right)+2\right)\\ \nonumber &&-f \dot{t}^2 X \epsilon  \frac{dP_{l}}{d\theta} \cos (\sigma  t)\bigg\},
\end{eqnarray}
\begin{eqnarray}
\label{cd3}
\dot{\phi}&&=\frac{L_{z}\csc^{2}\theta}{r^{2}[1+\epsilon(ZP_{l}+W\frac{dP_{l}}{d\theta}\cot\theta)\cos(\sigma t)]},
\end{eqnarray}
where the dots denote derivatives with respect to the proper time $\tau$, and the primes denote derivatives with respect to the radial coordinate $r$. It is obvious that the geodesic equations (\ref{cdx}-\ref{cd3}) are not variable-separable and the integration constants in this dynamical system are less than the number of freedom degrees. It implies that the photon dynamical system is non-integrable, so the chaos could appear in the motion of photon in this spacetime (\ref{dgmtr}).

\section{Shadows casted by Schwarzschild black hole perturbed by gravitational wave}

In this section, we will study the shadows of Schwarzschild black hole with the gravitational wave (\ref{dg}) through the backward ray-tracing technique \cite{sw,swo,astro,chaotic,binary,sha18,my,sMN,swo7,mbw}. We evolved light rays by solving numerically the null geodesic equations (\ref{cdx}-\ref{cd3}) from the observer backward in time. The shadow of black hole is composed by the light rays falling down into the horizon of black hole. We can find the spactime of Schwarzschild black hole with the gravitational wave is not asymptotically flat under the influence of gravitational perturbation (\ref{dg}). The observer can not be set at the spatial infinite. In this situation, we introduce orthonormal tetrads for observers located at finite distance, zero-angular-moment-observers (ZAMOs) reference frame \cite{sha2}, which has been used in the study of black hole shadow in de Sitter spacetime \cite{reisit,light,shads}. In this way, we assume that the static observer is locally at ($r_{obs}, \theta_{obs}$) in the Boyer-Lindquist coordinates.
The local observer basis  $\{e_{\hat{t}},e_{\hat{r}},e_{\hat{\theta}},e_{\hat{\phi}}\}$ can be expanded as a form in the coordinate basis $\{\partial_t,\partial_r,\partial_{\theta},\partial_{\phi} \}$
\cite{sw,swo,astro,chaotic,binary,sha18,my,sMN,swo7,mbw}
\begin{eqnarray}
\label{zbbh}
e_{\hat{\mu}}=e^{\nu}_{\hat{\mu}} \partial_{\nu},
\end{eqnarray}
where the transform matrix $e^{\nu}_{\hat{\mu}}$ obeys $(g_{\mu\nu}+\epsilon h_{\mu\nu})e^{\mu}_{\hat{\alpha}}e^{\nu}_{\hat{\beta}}
=\eta_{\hat{\alpha}\hat{\beta}}$, and $\eta_{\hat{\alpha}\hat{\beta}}$ is the Minkowski metric. For the spacetime (\ref{dg}), it is convenient to choice the transform matrix $e^{\nu}_{\hat{\mu}}$ as
\begin{eqnarray}
\label{zbbh1}
e^{\nu}_{\hat{\mu}}=\left(\begin{array}{cccc}
\frac{1}{\sqrt{-g_{tt}-\epsilon h_{tt}}}&0&0&0\\
0&\frac{1}{\sqrt{g_{rr}+\epsilon h_{rr}}}&0&0\\
0&0&\frac{1}{\sqrt{g_{\theta\theta}+\epsilon h_{\theta\theta}}}&0\\
0&0&0&\frac{1}{\sqrt{g_{\phi\phi}+\epsilon h_{\phi\phi}}}
\end{array}\right).
\end{eqnarray}
Thus the locally measured four-momentum $p^{\hat{\mu}}$ of a photon is related to its four-momentum $p^{\mu}$  by
 $p^{\hat{\mu}}=e^{\nu}_{\hat{\mu}} p_{\nu}$, i.e.,
\begin{eqnarray}
\label{dl}
p^{\hat{t}}&=&-e^{\nu}_{\hat{t}} p_{\nu}=-\frac{p_{t}}{\sqrt{f+\epsilon fXP_{l}\cos(\sigma t)}}, \nonumber\\
p^{\hat{r}}&=&e^{\nu}_{\hat{r}} p_{\nu}=\sqrt{\frac{f}{1+\epsilon YP_{l}\cos(\sigma t)}}p_{r} ,\nonumber\\
p^{\hat{\theta}}&=&e^{\nu}_{\hat{\theta}} p_{\nu}=\frac{p_{\theta}}{r\sqrt{1+\epsilon(ZP_{l}+W\frac{d^{2}P_{l}}{d\theta^{2}})\cos(\sigma t)}},
\nonumber\\
p^{\hat{\phi}}&=&e^{\nu}_{\hat{\phi}} p_{\nu}=\frac{L_z\csc\theta}{r\sqrt{1+\epsilon(ZP_{l}+W\frac{dP_{l}}{d\theta}\cot\theta)\cos(\sigma t)}}.
\end{eqnarray}
Repeating the operation in Refs.\cite{sw,swo,astro,chaotic,binary,sha18,my,sMN,swo7,mbw}, one can obtain the coordinates of a photon's image in observer's sky
\begin{eqnarray}
\label{xd1}
x&=&-r\frac{p^{\hat{\phi}}}{p^{\hat{r}}}|_{(r_{obs},\theta_{obs})}
=-\sqrt{\frac{f}{[1+\epsilon(ZP_{l}+W\frac{dP_{l}}{d\theta}\cot\theta)\cos(\sigma t)][1+\epsilon YP_{l}\cos(\sigma t)]}}\frac{L_{z}\csc\theta}{\dot{r}}|_{(r_{obs},\theta_{obs})}, \nonumber\\
y&=&r\frac{p^{\hat{\theta}}}{p^{\hat{r}}}|_{(r_{obs},\theta_{obs})}=
\sqrt{\frac{f[1+\epsilon(ZP_{l}+W\frac{d^{2}P_{l}}{d\theta^{2}})\cos(\sigma t)]}{1+\epsilon YP_{l}\cos(\sigma t)}}\frac{r^{2}\dot{\theta}}{\dot{r}}|_{(r_{obs},\theta_{obs})}.
\end{eqnarray}

\begin{figure}
\includegraphics[width=16.5cm ]{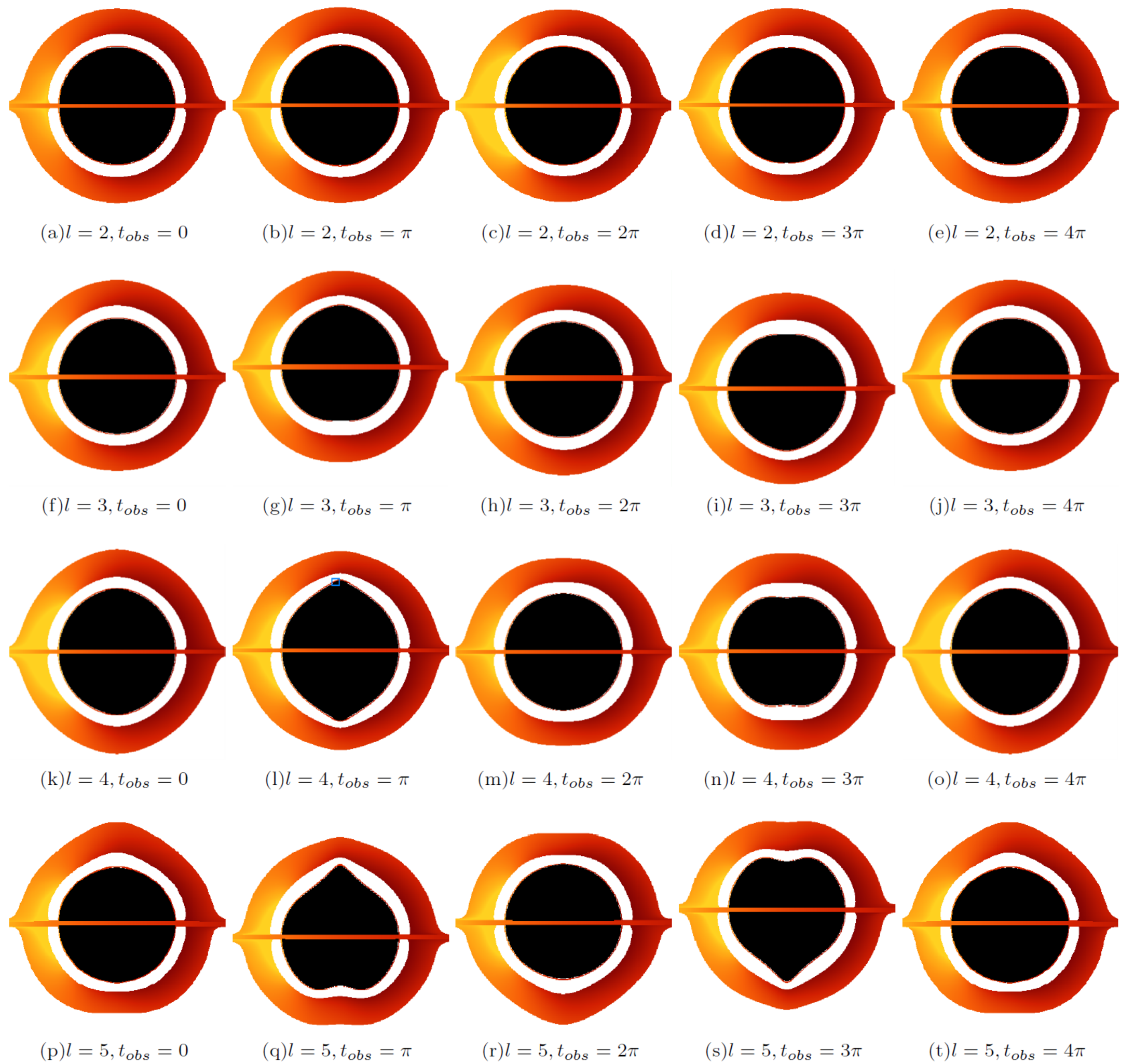}
\caption{The shadows of Schwarzschild black hole perturbed by the gravitational wave (\ref{dg}) at different times $t_{obs}$. Here we set mass $M=1$, $\epsilon=0.05$, gravitational wave frequency $\sigma=0.5$, $l=2,3,4,5$, and the observer is set at $r_{obs}=50$ with the inclination angle $\theta_{obs}=90^{\circ}$. In figure (l), the boundary of shadow in the region within the blue box is not smooth. We zoomed in on this region and got Fig.\ref{fx}(a) in which we found some more fine structures.}
\label{0.05-0.5}
\end{figure}

Here, we assume the lights come from both accretion disk around black hole and distant stars. For a convenience, we assume that a geometrically thin and optically thick accretion disk lies in the equatorial plane of Schwarzschild black hole. As that in ref.\cite{lum}, the disk's minimum and maximum radii are set as $6M$ and $15M$, respectively. In Fig.\ref{0.05-0.5}, we present the shadows of Schwarzschild black hole perturbed by the gravitational wave (\ref{dg}) for the static observer ($r_{obs}=50$, $\theta_{obs}=90^{\circ}$) at different times $t_{obs}$. Here we set black hole mass $M=1$, the parameter $\epsilon=0.05$, gravitational wave frequency $\sigma=0.5$, $l=2,3,4$ and $5$. Obviously, the black regions represent shadows of black hole, the colored regions (from red to yellow) represent the images of accretion disk. In addition, because of the rotation of accretion disk, the redshift of photon changes the colors in accretion disk images from red to yellow,  which depends on the flux radiation of accretion disk. In this case, the yellow has bigger flux radiation. From this figure, we can find the shadows of Schwarzschild black hole perturbed by gravitational wave change periodically with time, which can be understand by a fact that the metric coefficients of the black hole (\ref{dg}) are the periodic functions of time $t$. Moreover, the shadows also depend on the parameters of gravitational wave. In Fig.\ref{0.05-0.5}, we find that the black hole shadow is symmetric with respect equatorial plane in the even $l$ case, but they are not in the odd $l$ case. It means that the symmetry of black hole shadow also depends on the symmetry of the gravitational perturbation (\ref{dg}). For the case of the perturbation with odd $l$, it is found that the center of black hole shadow oscillates up and down with time along the direction which is vertical to equatorial plane. However, for the even $l$ one, the center of the shadow doesn't change, but black hole shadow alternately stretches and squeezes with time along the vertical direction. Moreover, the shape of black hole shadows changes only along the vertical direction in Fig.\ref{0.05-0.5}. It is because that the special class of gravitational wave (\ref{dg}) is only an axisymmetric perturbation.  With the increase of the parameter $l$, we find that the deformations of shadows are more obvious. Especially when $l$ increases to $5$, the shadows of black hole even become heart-like shaped shadows. In addition, we also find that the images of accretion disk have the same variations as the black hole shadows with gravitational waves.

In order to study quantitatively the deformation of Schwarzschild black hole shadow perturbed by the gravitational perturbation (\ref{dg}), we define two deviated parameters $\varepsilon_{o}$ and $\varepsilon_{e}$, which can describe the deviation strength of the shadows from usual Schwarzschild black hole in the odd $l$ case and in the even $l$ case, respectively. Firstly, we must introduce four important points for shadow, the leftmost point ($x_{l}$, $y_{l}$), the rightmost point ($x_{r}$, $y_{r}$), the topmost point ($x_{t}$, $y_{t}$) and the bottommost point ($x_{b}$, $y_{b}$), where the coordinates ($x$, $y$) are the celestial coordinates (\ref{xd1}) in observer's sky. In the odd $l$ case, the center of shadow oscillates up and down with time along the vertical direction, and we found the leftmost point ($x_{l}$, $y_{l}$) and the rightmost point ($x_{r}$, $y_{r}$) oscillate  with the shadow center, so the deviated parameter $\varepsilon_{o}$ could be defined as
\begin{eqnarray}
\label{d3}
\varepsilon_{o}=y_{l}(y_{r}).
\end{eqnarray}
The deviated parameter $\varepsilon_{o}$ is positive if black hole shadow shifts upward, but is negative if the shadow shifts downward. Since the shadow in the even $l$ case alternately stretches and squeezes along the vertical direction, and then the deviated parameter $\varepsilon_{e}$ could be defined as
\begin{eqnarray}
\label{d4}
\varepsilon_{e}=\frac{y_{t}}{x_{r}}-1.
\end{eqnarray}
\begin{figure}
\includegraphics[width=15cm ]{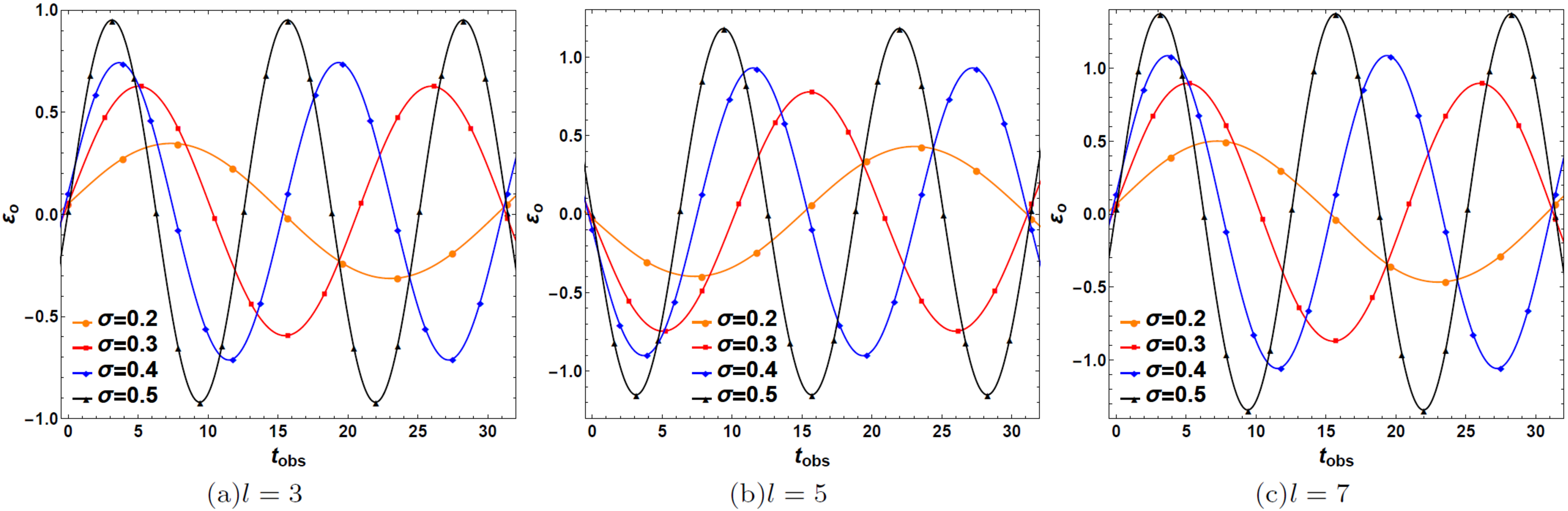}
\caption{The changes of the deviated parameter $\varepsilon_{o}$ with time $t_{obs}$. Here we set mass $M=1$, $\epsilon=0.05$, gravitational wave frequency $\sigma=0.2,0.3,0.4,0.5$, and $l=3,5,7$.}
\label{wsj}
\end{figure}
\begin{figure}
\includegraphics[width=15cm ]{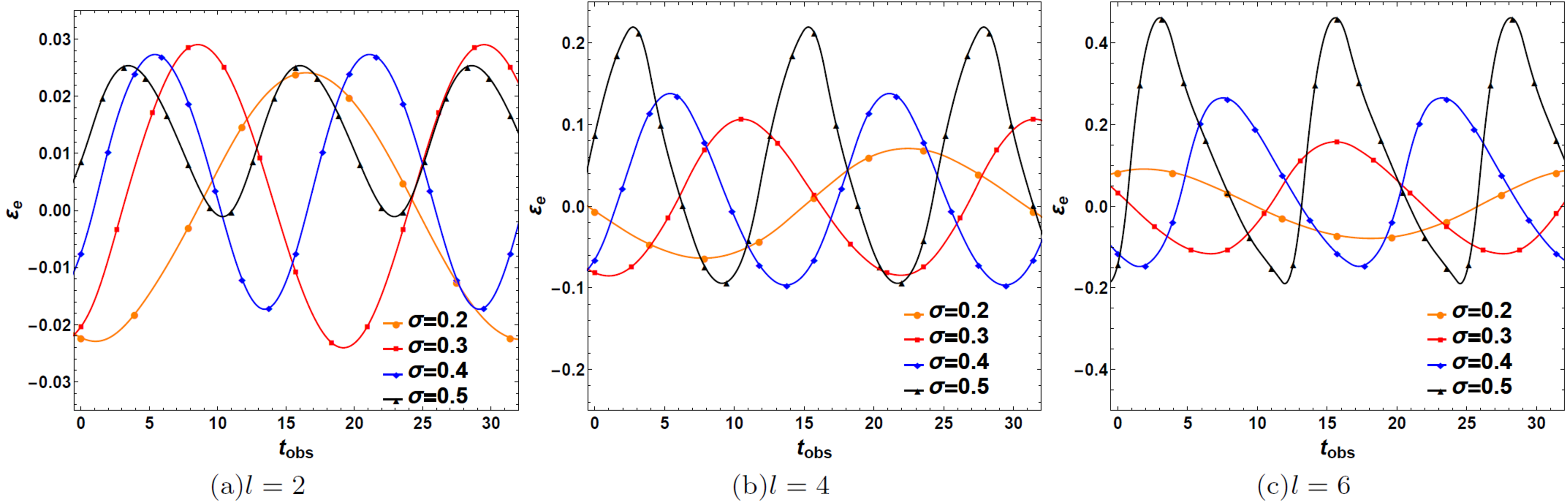}
\caption{The changes of the deviated parameter $\varepsilon_{e}$ with time $t_{obs}$. Here we set mass $M=1$, $\epsilon=0.05$, gravitational wave frequency $\sigma=0.2,0.3,0.4,0.5$, and $l=2,4,6$.}
\label{ysj}
\end{figure}
The deviated parameter $\varepsilon_{e}$ is positive if black hole shadow is prolate, but is negative if the shadow is oblate.
Figs.\ref{wsj} and \ref{ysj} show that the deviated parameters $\varepsilon_{o}$ for $l=3, 5, 7$ and $\varepsilon_{e}$ for $l=2, 4, 6$ both fluctuate up and down with time, which indicates once again that the shadow perturbed by the gravitational perturbation (\ref{dg}) oscillates up and down with time along the vertical direction in the odd $l$ case, and alternately stretches and squeezes with time along the vertical direction in the even $l$ case. In Fig.\ref{ysj}(a), from the amplitude values of deviated parameter $\varepsilon_{e}$ one can find the black hole shadows change very little for $l=2$, and the amplitude of deviated parameter $\varepsilon_{e}$ decreases with frequency $\sigma$ increases. But for $l=3, 4, 5, 6$ and $7$, one can find the amplitudes of deviated parameters $\varepsilon_{e}$ and $\varepsilon_{o}$ both increase with the frequency $\sigma$ increases. What's more, the deviated parameter $\varepsilon_{e}$ is greater than zero for a little longer in Fig.\ref{ysj}, which indicates that the gravitational perturbation (\ref{dg}) make Schwarzschild black hole shadow more prolate rather than oblate when $l$ is even. In addition, we found that both of the amplitude of  $\varepsilon_{o}$ and $\varepsilon_{e}$ increase with $l$ for the same frequency $\sigma$. It means that the bigger $l$ has a greater impact on the black hole shadow.

\begin{figure}
\includegraphics[width=16cm ]{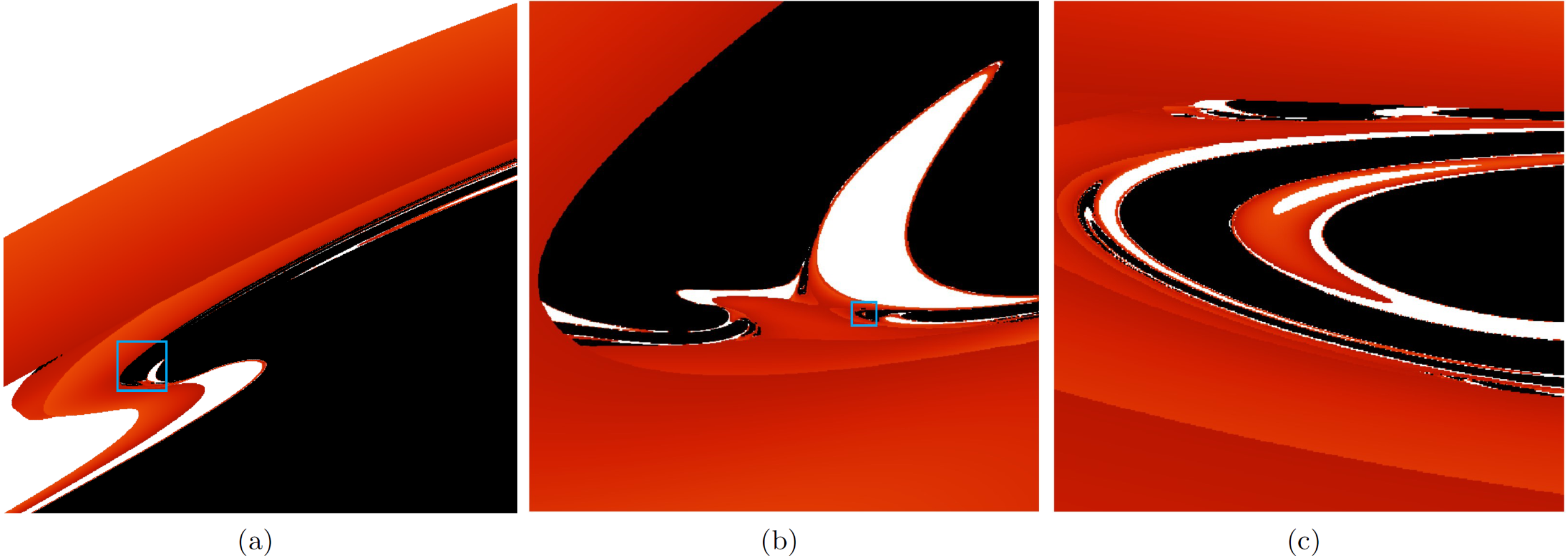}
\caption{(a)The amplifying image of the area within the blue box in Fig.\ref{0.05-0.5}(l). (b)The amplifying image of the area within the blue box in (a). (c)The amplifying image of the area within the blue box in (b).}
\label{fx}
\end{figure}

In Schwarzschild black hole spacetime, the boundary of shadow is a smooth curve. However, in Fig.\ref{0.05-0.5}(l), one can find that the boundary of shadow in the blue box is not smooth. We amplified the boundary of shadow and found some similar layered structures shown in Fig.\ref{fx}(a). From this figure we can find the boundary of shadow is distorted, the color red and white near the shadow represent the lights come from accretion disk and distant stars respectively. They're superimposed layer by layer. We zoomed in on the region within the blue box in Fig.\ref{fx}(a) to got Fig.\ref{fx}(b), and continued to zoom in on the region within the blue box in Fig.\ref{fx}(b) to got Fig.\ref{fx}(c). We kept amplifying the boundary of shadow and found more similar layered structures like shown in Fig.\ref{fx}. Actually, they are self-similar fractal structures caused by the chaotic motion of photon in the background of a Schwarzschild black hole perturbed by the gravitational wave. At arbitrarily small scale, this shadow boundary region always has finer self-similar structures. It's so irregular that can't be described by calculus or traditional geometric language. The new effect that this special kind of gravitational wave (\ref{dg}) gives to Schwarzschild black hole shadow is what we expected since the equations of photon motion (\ref{cdx}-\ref{cd3}) are no longer integrable and the chaotic motions for test particles have appeared \cite{cgw} in this background. I hope Event Horizon Telescope and BlackHoleCam could find more fine structure like chaos by increasing the resolution of telescope in the future.

\begin{figure}
\includegraphics[width=16.5cm ]{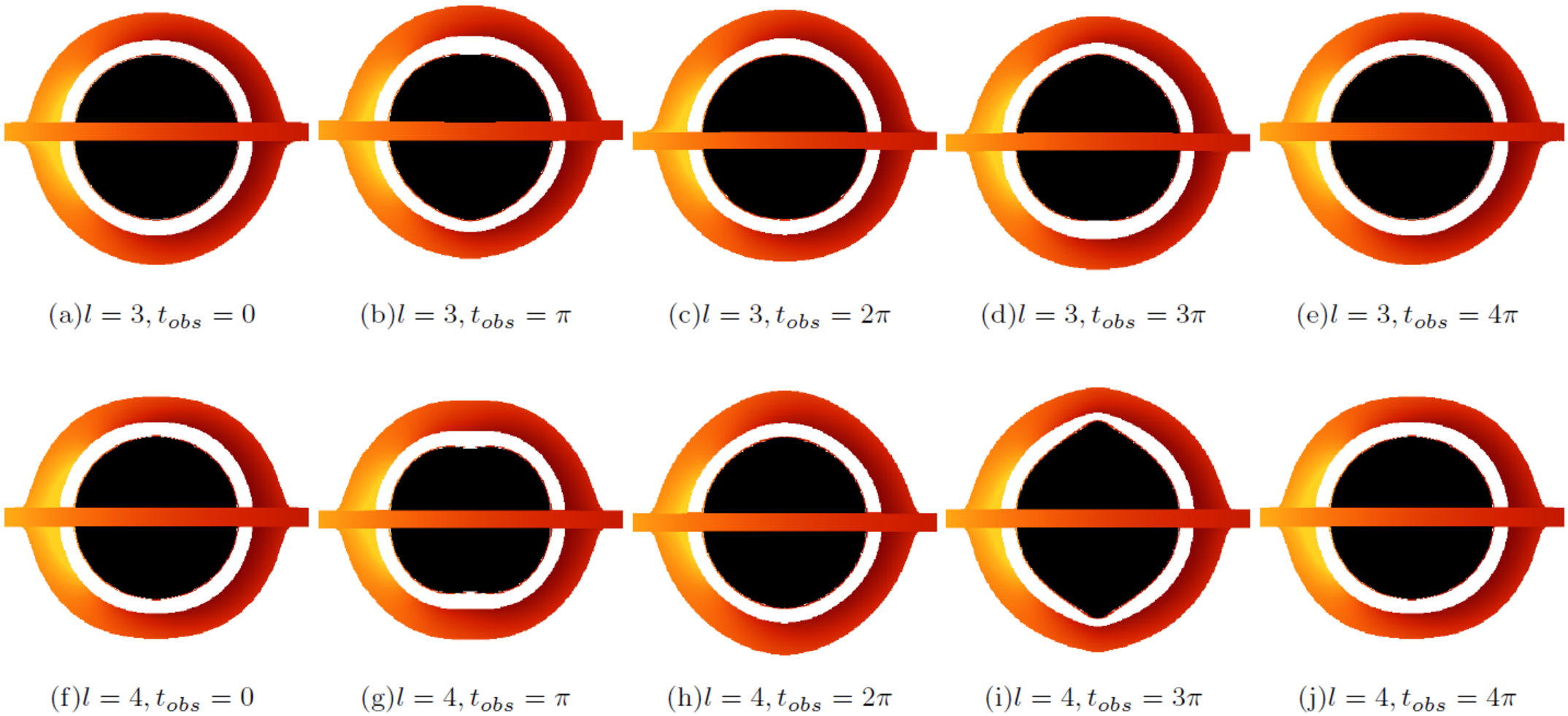}
\caption{The shadows of Schwarzschild black hole perturbed by the gravitational wave (\ref{dg}) at different times $t_{obs}$. Here we set mass $M=1$, $\epsilon=0.05$, gravitational wave frequency $\sigma=0.5$, $l=3,4$, and the observer is set at $r_{obs}=20$ with the inclination angle $\theta_{obs}=90^{\circ}$.}
\label{r20}
\end{figure}

\begin{figure}
\includegraphics[width=16.5cm ]{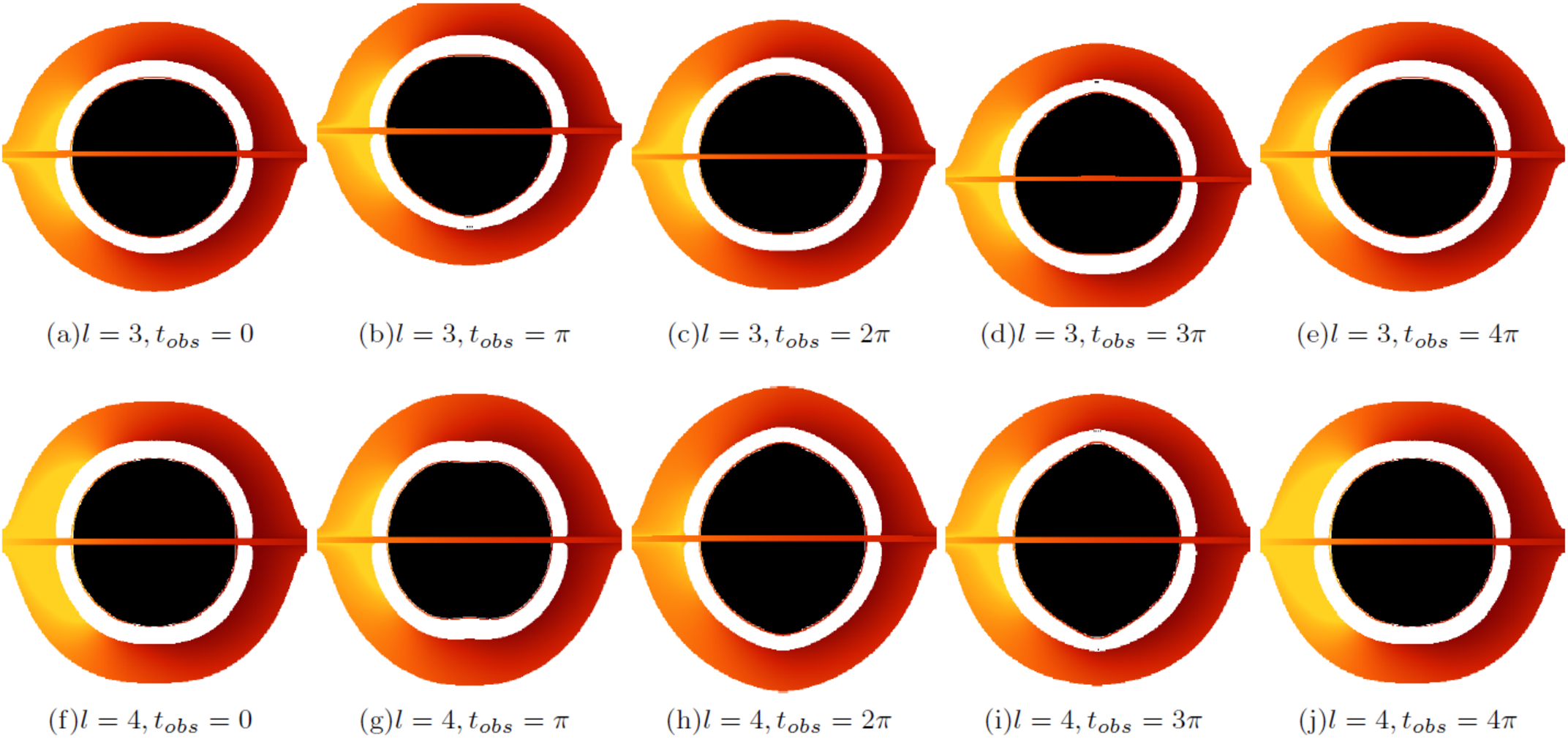}
\caption{The shadows of Schwarzschild black hole perturbed by the gravitational wave (\ref{dg}) at different times $t_{obs}$. Here we set mass $M=1$, $\epsilon=0.05$, gravitational wave frequency $\sigma=0.5$, $l=3,4$, and the observer is set at $r_{obs}=80$ with the inclination angle $\theta_{obs}=90^{\circ}$.}
\label{r80}
\end{figure}

\begin{figure}
\includegraphics[width=13cm ]{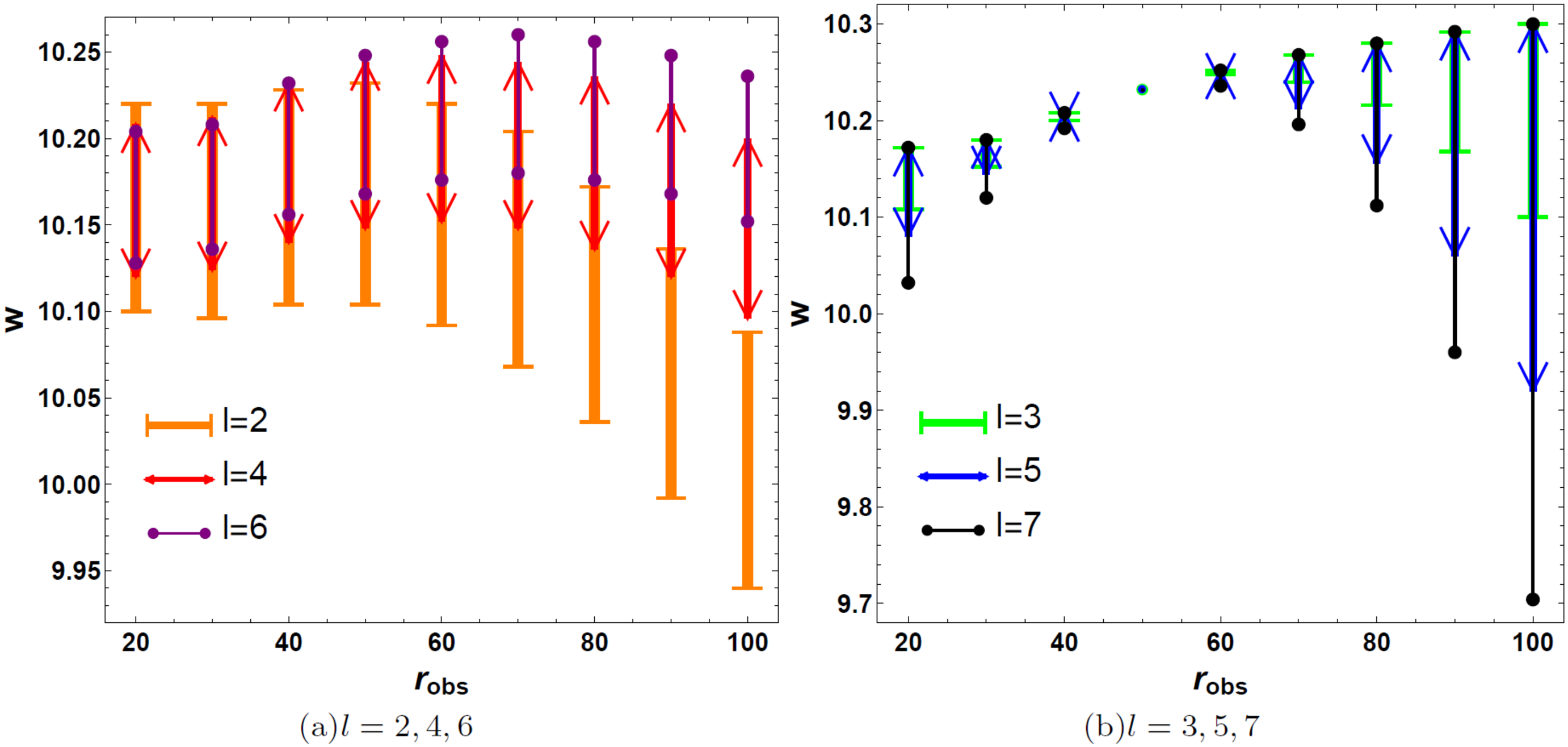}
\caption{The width $w=x_{r}-x_{l}$ of the shadow of Schwarzschild black hole perturbed by gravitational wave (\ref{dg}) for the static observer at different radial coordinate $r_{obs}$. The length of the lines represents the variation range of the width $w$ of black hole shadow in a time-period. We set $M=1$, $\epsilon=0.05$, gravitational wave frequency $\sigma=0.5$, $l=2,4,6$ for (a) and $l=3,5,7$ for (b).}
\label{w}
\end{figure}

\begin{figure}
\includegraphics[width=13cm ]{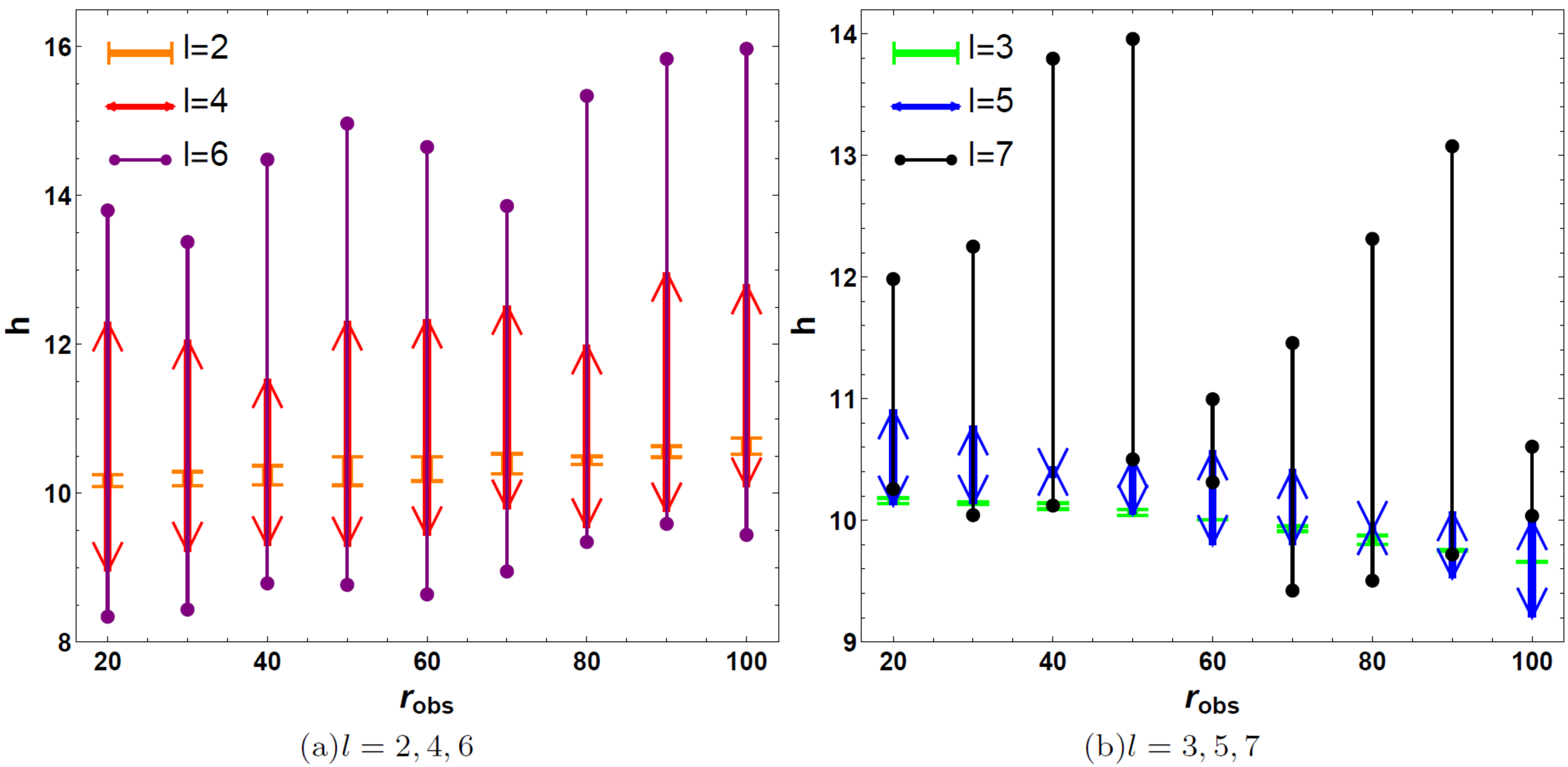}
\caption{The height $h=y_{t}-y_{b}$ of the shadow of Schwarzschild black hole perturbed by gravitational wave (\ref{dg}) for the static observer at different radial coordinate $r_{obs}$. The length of the lines represents the variation range of the height $h$ of black hole shadow in a time-period. We set $M=1$, $\epsilon=0.05$, gravitational wave frequency $\sigma=0.5$, $l=2,4,6$ for (a) and $l=3,5,7$ for (b).}
\label{h}
\end{figure}

\begin{figure}
\includegraphics[width=13cm ]{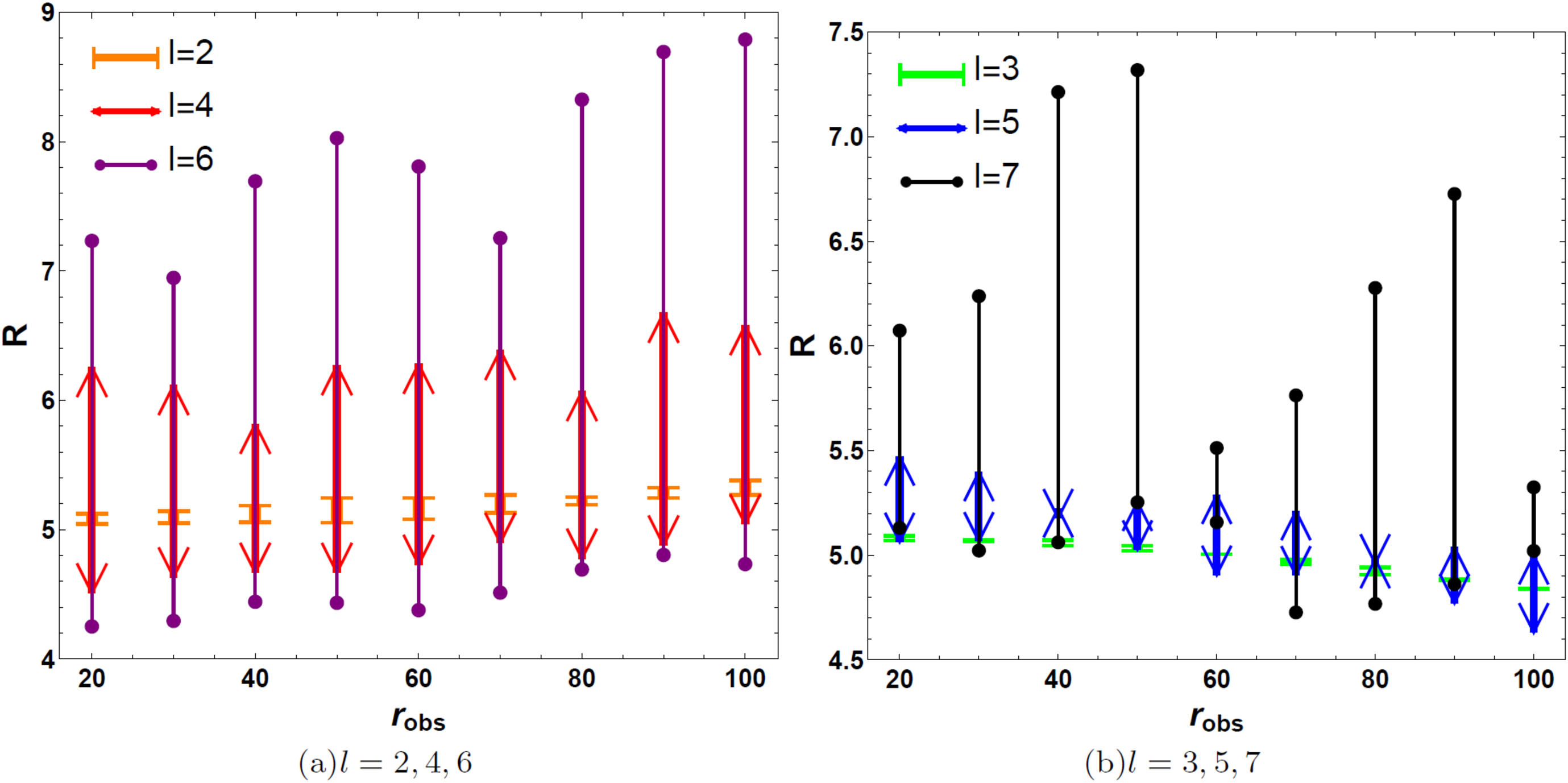}
\caption{The radius $R$ of the shadow of Schwarzschild black hole perturbed by gravitational wave (\ref{dg}) for the static observer at different radial coordinate $r_{obs}$. The length of the lines represents the variation range of the radius $R$ of black hole shadow in a time-period. We set $M=1$, $\epsilon=0.05$, gravitational wave frequency $\sigma=0.5$, $l=2,4,6$ for (a) and $l=3,5,7$ for (b).}
\label{R}
\end{figure}

Let us now to probe the changes of the black hole shadow with the spatial coordinate of observer.
In Fig.\ref{r20} and Fig.\ref{r80}, we present the shadows perturbed by gravitational wave (\ref{dg}) for the static observer at $r_{obs}=20$ and $80$, respectively. For a sake of simplicity, we set $\theta_{obs}=90^{\circ}$, $M=1$, $\epsilon=0.05$, and gravitational wave frequency $\sigma=0.5$, $l=3,4$. Comparing Fig.\ref{0.05-0.5} with Fig.\ref{r20} and Fig.\ref{r80}, one can find with the increase of $r_{obs}$, the image of accretion disk on the equatorial plane becomes thin, but the disk images around shadow become thick. The oscillation amplitude of black hole shadow increases with the coordinate $r_{obs}$ when $l$ is odd. Moreover, we also present the width $w=x_{r}-x_{l}$, the height $h=y_{t}-y_{b}$ and the radius $R$ of the black hole shadow for the static observer with different radial coordinates in Figs.\ref{w}-\ref{R}, respectively. Here the radius $R$ is defined similarly as in Ref.\cite{sha9}
\begin{eqnarray}
\label{rs}
R=\frac{x_{r}^{2}+(\frac{y_{t}}{2}-\frac{y_{b}}{2})^{2}}{2x_{r}}.
\end{eqnarray}
In these figures, we set mass $M=1$, $\epsilon=0.05$, gravitational wave frequency $\sigma=0.5$, and $l=2,4,6$ for figure (a), $l=3,5,7$ for figure (b). The length of the lines represents the variation range of the width $w$, the height $h$ or the radius $R$ of black hole shadow in a time-period. In Fig.\ref{w}, one can find the variation range of the width $w$ in a time-period is bigger for small even $l$, but it is just the opposite for odd $l$. Especially, the widths $w$ almost unchange when $r_{obs}=50$ for $l=3,5,7$. Moreover, one can find the average width $w$ in a time-period increases first and then decreases with $r_{obs}$. Unlike the width $w$ of black hole shadow, one can find the variation range of the height $h$ in a time-period changes periodically with $r_{obs}$ in Fig.\ref{h} and the corresponding period is about $40$. It is very interesting and worthy of further study.
Moreover, one can found the variation range of the height $h$ in a time-period is larger for bigger $l$. Overall the average height $h$ in a time-period increases for even $l$ and decreases for odd $l$ with $r_{obs}$. From Fig.\ref{w} and \ref{h} one can find that the change of the black hole shadow height $h$ is bigger than that of the shadow width $w$, it means that the gravitational perturbation (\ref{dg}) has a greater influence on the vertical direction of black hole shadow. So in Fig.\ref{R} we can find the change of the radius $R$ of black hole shadow is similar to that of the black hole shadow height $h$.

\begin{figure}
\includegraphics[width=16.5cm ]{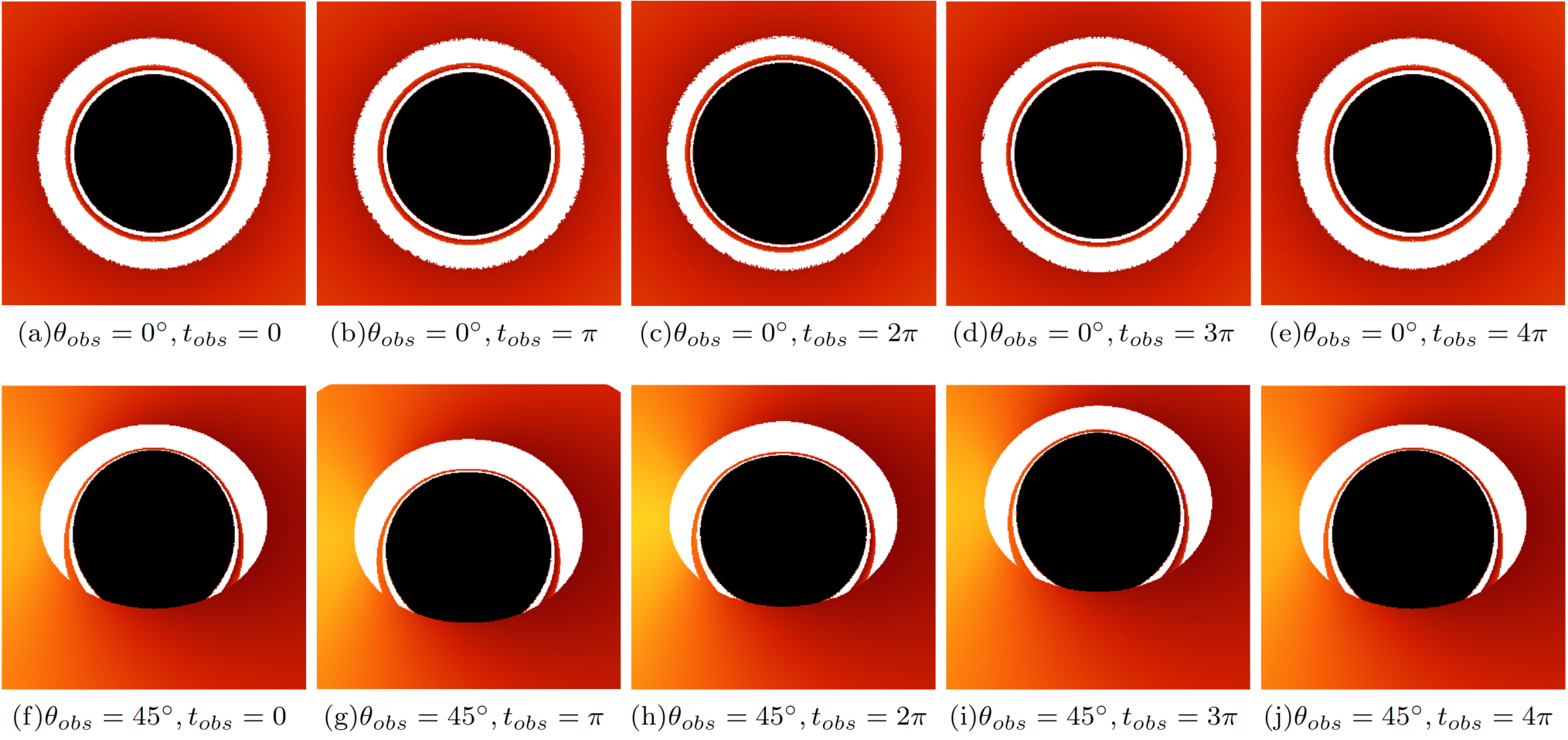}
\caption{The shadows for the observers at inclination angle $\theta_{obs}=0^{\circ}$ (upper row) and $45^{\circ}$ (bottom row) in different times $t_{obs}$. Here we set $r_{obs}=50$, $M=1$, $\epsilon=0.05$, gravitational wave frequency $\sigma=0.5$, $l=3$.}
\label{l3}
\end{figure}

\begin{figure}
\includegraphics[width=16.5cm ]{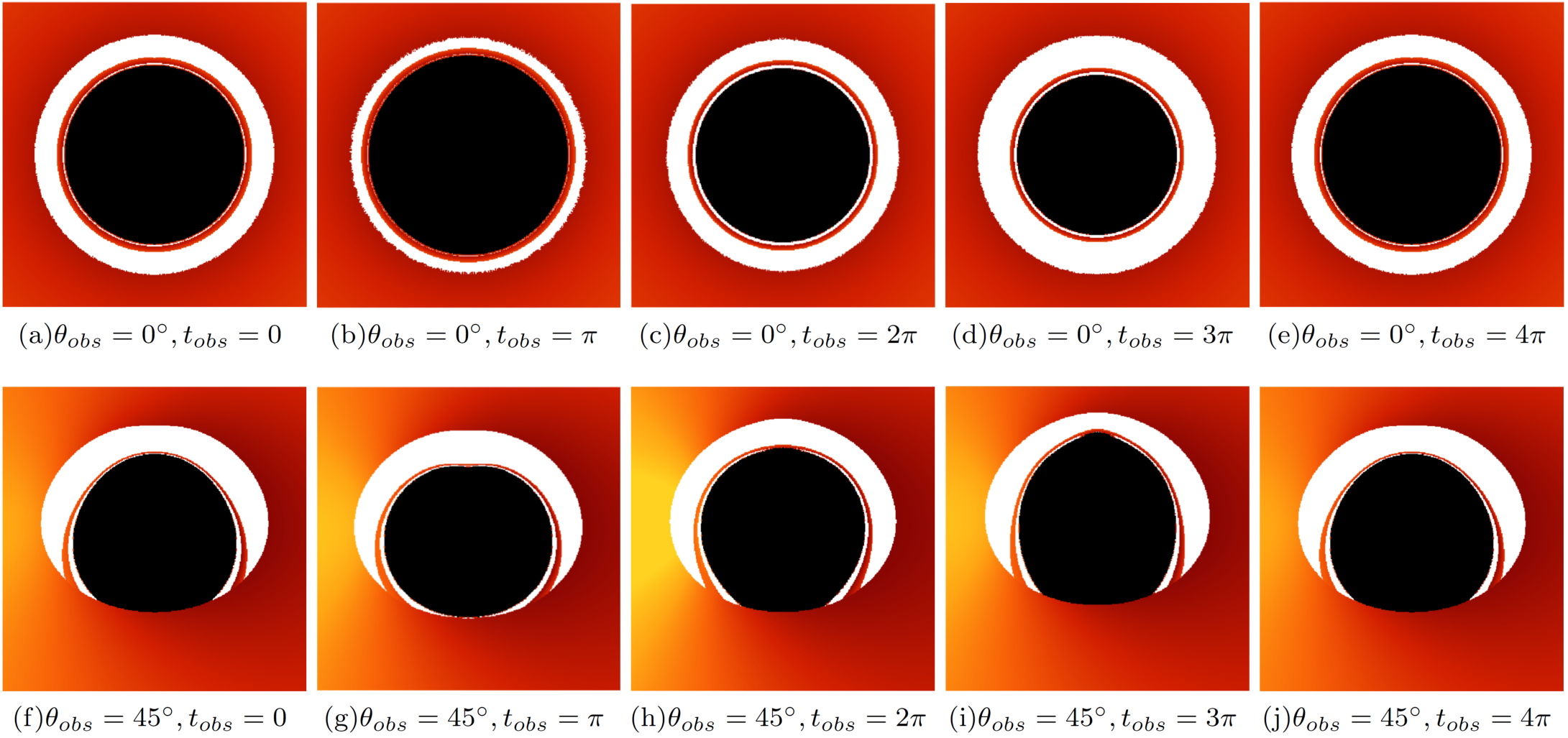}
\caption{The shadows for the observers at inclination angle $\theta_{obs}=0^{\circ}$ (upper row) and $45^{\circ}$ (bottom row) in different times $t_{obs}$. Here we set $r_{obs}=50$, $M=1$, $\epsilon=0.05$, gravitational wave frequency $\sigma=0.5$, $l=4$.}
\label{l4}
\end{figure}

\begin{figure}
\includegraphics[width=16cm ]{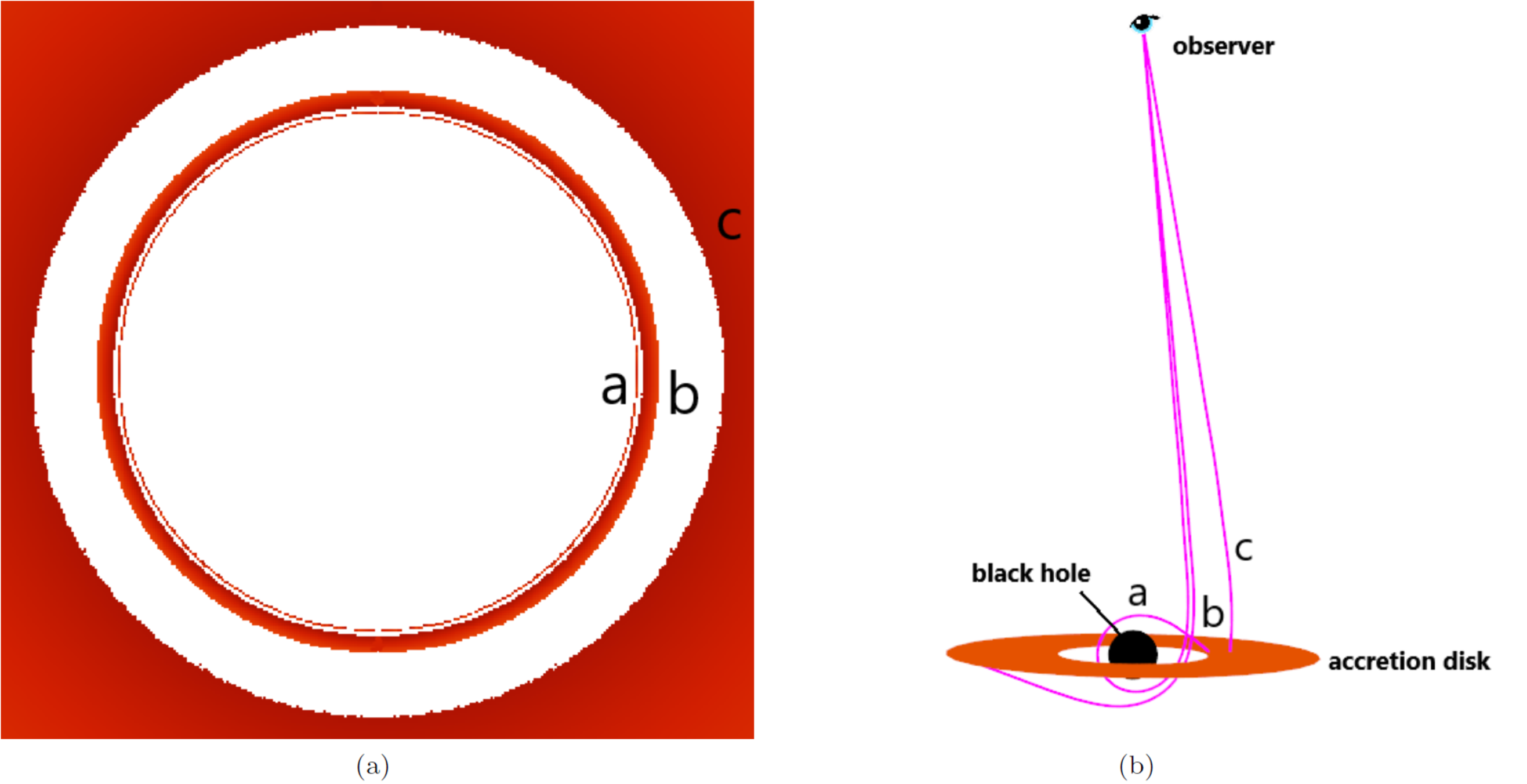}
\caption{(a)The three peaks of emitted intensity of light rays in Fig.\ref{l4}(a), the ring $a$, the thin image $b$ and the region $c$. (b)The three light rays from accretion disk to observer, which correspond to the ring $a$, the thin image $b$ and the region $c$ in (a).}
\label{gj}
\end{figure}

In Fig.\ref{l3} and \ref{l4}, we show the dependence of the shadows on the observer inclination angle $\theta_{obs}$ for $l=3$ and $4$ respectively. When $\theta_{obs}=0^{\circ}$ (upper row), one can find that the black hole shadow alternately expands and contracts with time under the influence of gravitational waves(\ref{dg}). When $\theta_{obs}=45^{\circ}$ (bottom row), the shadow is no longer symmetric with respect the equatorial plane. But we also find the shadow of black hole oscillates up and down with time for $l=3$, and the shadow alternately stretches and squeezes with time along the vertical direction for $l=4$. The colored region (red to yellow) also represents the image of accretion disk. Especially, in these figures one can find the emitted intensity of light rays have three peaks, one is close to the boundary of black hole shadow, one is the middle thin accretion disk image, and one is the peripheral accretion disk image. For instance, the three peaks of emitted intensity of light rays in Fig.\ref{l4}(a) are shown in Fig.\ref{gj}(a), which are the ring $a$, the thin image $b$ and the region $c$ respectively. Through further study, we found the light rays from the region $c$ come from the upside of accretion disk and propagate directly to observer, as the trajectory $c$ in Fig.\ref{gj}(b). The light rays from the thin image $b$ come from the downside of accretion disk and bypass the black hole to observer, as the trajectory $b$ in Fig.\ref{gj}(b). The light rays from the ring $a$ also come from the upside of accretion disk, but it make one orbit around black hole and then reaching the observer, as the trajectory $a$ in Fig.\ref{gj}(b). Actually, when the metric (\ref{dgmtr}) reduces to Schwarzschild black hole (without the gravitational perturbation, $\epsilon=0$), the three peaks of emitted intensity of light rays still exist. It is because the accretion disk has inner and outer boundaries, and the lights can propagate between black hole and accretion disk, which causes the multiple images of the accretion disk to appear. In addition, the image of accretion disk has the same variations as black hole shadow, it alternately expands and contracts with time when $\theta_{obs}=0^{\circ}$, and is no longer symmetric with respect the equatorial plane when $\theta_{obs}=45^{\circ}$, but also oscillates up and down with time for $l=3$, alternately stretches and squeezes with time along the vertical direction for $l=4$.

\section{Summary}
In this paper we studied the shadows of a Schwarzschild black hole perturbed by a special class of gravitational wave (\ref{dg}). Under the influence of gravitational wave, the equations of photon motion in this spacetime are no longer integrable, and the shadow changes periodically. When the order of Legendre polynomial $l$ is odd, the center of shadow oscillates up and down with time along the direction which is vertical to equatorial plane. When $l$ is even, the center of shadow does not move, but it alternately stretches and squeezes with time along the vertical direction. We studied the effects of gravitational wave on Schwarzschild black hole shadows by introducing two deviated parameters $\varepsilon_{o}$ and $\varepsilon_{e}$, and found both of the amplitude of $\varepsilon_{o}$ and $\varepsilon_{e}$ increase with $l$ increases for fixed $\sigma$. We also found that there exist self-similar fractal structures in the boundary of shadow caused by chaotic motion of photon due to the presence of gravitational wave. The average width $w$ of black hole shadow in a time-period increases first and then decreases overall as $r_{obs}$ increases. The variation range of height $h$ and radius $R$ of black hole shadow change periodically with $r_{obs}$. Moreover, the change of the black hole shadow height $h$ is bigger than that of the shadow width $w$, which indicates the gravitational perturbation (\ref{dg}) has a greater influence on the vertical direction of black hole shadow. In addition, we present the shadows for the observer with different inclination angle. Our results show that the presence of gravitational wave yields some interesting properties of black hole shadow.

\section{\bf Acknowledgments}

This work was supported by the Shandong Provincial Natural Science Foundation of China under Grant No. ZR2020QA080, and was partially supported by the National Natural Science Foundation of China under Grant No. 11875026, 11475061, 11875025, and 12035005.

\end{document}